\documentclass[12pt]{article}
\usepackage{pdproc,epsfig} 

\begin{document}
\def\fluxunit{cm$^{-2}$s$^{-1}$MeV$^{-1}$}
\def\sfrunit{M$_\odot$yr$^{-1}$Mpc$^{-3}$}
\def\snrate{yr$^{-1}$Mpc$^{-3}$}
\def\beq{\begin{equation}}
\def\eeq{\end{equation}}

\def\ga{\mathrel{\raise.3ex\hbox{$>$\kern-.75em\lower1ex\hbox{$\sim$}}}}
\def\la{\mathrel{\raise.3ex\hbox{$<$\kern-.75em\lower1ex\hbox{$\sim$}}}}
\renewcommand{\thefootnote}{\alph{footnote}}

\title{
  THE RELIC NEUTRINO BACKROUND FROM THE FIRST STARS\footnote{Summary of talk given at the IIIrd International Workshop on:
NO-VE  "Neutrino Oscillations in Venice", Venice Italy, February 2006}}

\author{KEITH A. OLIVE and PEARL SANDICK}

\address{ William I. Fine Theoretical Physics Institute, University of Minnesota\\
Minneapolis, MN 55416, USA\\
 {\rm E-mail: olive@umn.edu}}

%  \centerline{\footnotesize and}

%\author{SECOND AUTHOR'S NAME}

%\address{Group, Company, Address, City, State ZIP/Zone,Country}

\abstract{
We consider
the relic neutrino background produced by Population III
stars coupled with a normal mode of star formation at lower redshift.
The computation is performed in the framework of 
hierarchical structure formation 
and is based
on cosmic star formation histories constrained to reproduce the observed star formation
rate at redshift $z \la 6$, the observed chemical abundances in damped Lyman alpha absorbers
and in the intergalactic medium, and to allow for an early reionization of the Universe
 at $z\sim 10-20$. We consider both a burst and non-burst model for Population III star formation. 
We find that although the high redshift burst of Population III stars does
lead to an appreciable flux of neutrinos at relatively low energy ($E_\nu \approx 1$ MeV),  the observable neutrino flux is dominated by the normal mode of star formation.  We also find that predicted
fluxes are at the present level of the SuperK limit.  As a consequence, the supernova relic neutrino
background has a direct impact on models of chemical evolution and/or supernova dynamics. }
   
\normalsize\baselineskip=15pt

\section{Introduction}
\vspace*{-5.6in}
\rightline{hep-ph/0603236}
\rightline{UMN--TH--2436/06}
\rightline{FTPI--MINN--06/08}
\rightline{March 2006}
\vskip 5.0in
 Among the results contained in the first year data obtained by WMAP\cite{wmap}
was large optical depth implying  that the Universe became reionized at high redshift in the range, $11<z<30$ at 95\% CL. This can be explained by a generation of massive stars 
formed at high redshift\cite{reion}. Such a scenario would have notable consequences for 
cosmic chemical evolution, in particular,  the metal enrichment of the interstellar medium (ISM) and intergalactic medium (IGM)\cite{enrich,daigne1,daigne2}.  In addition, the cosmic star formation rate (SFR) at redshifts $z \la 6$ is observed to be over an order of magnitude larger
than the current SFR and is seen to have peaked at redshift $z \approx 3$~\cite{csfr}.

A direct consequence of this is the predicted enhancement
in the rate of core collapse supernovae.  In addition to the sharp spike of supernovae at very high redshift due to the explosions of stars responsible for the early epoch of reionization, the enhanced 
SFR of the normal mode of star formation at redshifts $z \la 6$ leads to a supernova rate
which is approximately a factor of 30 times the current rate, and a factor of 5 times the
observed rate at $z\sim 0.7$~\cite{dahlen}.

Another consequence of an enhanced SFR and SN rate is the resultant neutrino background spectrum
produced by the accumulated core collapse supernovae\cite{tot,ando,as,sksw,immrs,ourpaper}.  Here, we incorporate fully developed chemical evolution models which trace the history of
pre-galactic structures and are based on a $\Lambda$CDM cosmology and
include a Press-Schechter model\cite{ps} of hierarchical structure formation.
This will allow us to make detailed estimates of the predicted fluxes of neutrinos
which can be detected in large underground detectors.

\section{Chemical Evolution}

The basic elements of any chemical evolution model include:
the star formation rate, $\psi(t)$, which is observed to be greatly enhanced
at high redshift; the initial mass function, $\phi(m)$, assumed to have a near Salpeter slope
of 1.3;  element yields to allow us to compare with the chemical abundances in
DLAs, the IGM and metal-poor stars.  

The model employed is an open-box type model, which includes infall
due to the accretion of baryons in the hierarchal scenario, and outflows of 
gas due to supernova winds. 
The cosmic star formation histories we consider here have been adopted from detailed chemical
evolution models\cite{daigne1,daigne2}.  These models are bimodal and are described by a 
birthrate function of the form
\begin{equation}
B(m,t,Z) = \phi_1(m) \psi_1(t) + \phi_2(m) \psi_2(Z)
\label{birth}
\end{equation}
where $\phi_{1(2)}$ is the IMF of the normal (massive) component of star formation, and 
$\psi_{1(2)}$ is the respective star formation rate. $Z$ is the metallicity.
The normal component contains stars with masses between 0.1 and 100 M$_\odot$
and is primarily constrained by observations at low redshift ($z \la 6$). 
The massive component operates  at high redshift and is cut off once the metallicity reaches
a critical value taken to be $10^{-4}$. 
Both components can contribute to the chemical enrichment of galaxy forming structures and the IGM.
We examine three different models of the massive mode as described below.

We consider several minimal masses for minihalos for star formation to occur: $10^6, 10^7, 10^8, 10^9$, and $10^{11}$
M$_\odot$.  Star formation is assumed to begin when the baryon fraction
in the structures reaches $f_b = 0.01$.  This criteria fixes the initial redshift for star formation to begin.
For example, for $M_{\rm min} = 10^7$ M$_\odot$, star formation begins at $z = 16$. For
larger (smaller) masses, star formation begins later (earlier).
Later we will relax this assumption.

\section{The normal mode}

The normal mode of star formation is referred to as Model 0 and provides a standard star formation history, with stellar masses in the range 0.1 M$_{\odot} \le m \le 100$ M$_{\odot}$. 
The SFR for Model 0 is of an exponential form
\beq
\psi_1=\nu_1e^{-t/\tau_1},
\eeq 
which corresponds to a SFR dominated by elliptical galaxies.  
Best fits for $\nu_1$ and $\tau_1$ are given in Table 1.
In these models the outflow is non-zero and the details for computing the outflow are given in \cite{daigne1,daigne2}. The overall efficiency of outflow is parameterized by $\epsilon$ whose
value is also given in Table 1.
This model alone is inadequate for high redshift reionization.

\begin{table}[ht]
\begin{center}
\begin{tabular}{|ccccc|}
\multicolumn{5}{c}{\textbf{Normal mode}}  \\ 
\hline
$M_{\mathrm{min}}$ & 
$z_{\mathrm{init}}$ & 
$\epsilon$ & 
$\nu_{1}$  & 
$\tau_{1}$ \\
(M$_{\odot}$) & 
 & 
 & 
(Gyr$^{-1}$) & 
(Gyr)  \\
\hline
\hline
$10^{6}$  & 18.2 & $2\times 10^{-3}$ & 0.2 & 2.8 \\
\hline
$10^{7}$  & 16.0 & $3\times 10^{-3}$ & 0.2 & 2.8 \\
\hline
$10^{8}$  & 13.7 & $5\times 10^{-3}$ & 0.2 & 2.8 \\ 
\hline
$10^{9}$  & 11.3 & $        10^{-2}$ & 0.2 & 3.0 \\ 
\hline
$10^{11}$ & 6.57 & $1.5\times 10^{-2}$ & 0.5 & 2.2  \\ 
\hline
\end{tabular}
\end{center}
\caption{The model parameters for the normal mode of star formation (Model 0). Column 1 indicates the input value of the minimum mass for star forming structures. Column 2 is derived from column 1, having assumed that $f_b = 1\%$ when star formation begins.  In columns 3, 4, and 5, parameter values for the efficiency of outflow and the SFR are given.  The slope of the IMF is $x=1.3$ for all models. 
}
\label{tab:model0}
\end{table}

\begin{figure}
%\vspace*{13pt}
%\leftline{\hfill\vbox{\hrule width 5cm height0.001pt}\hfill}
      \mbox{\epsfig{figure=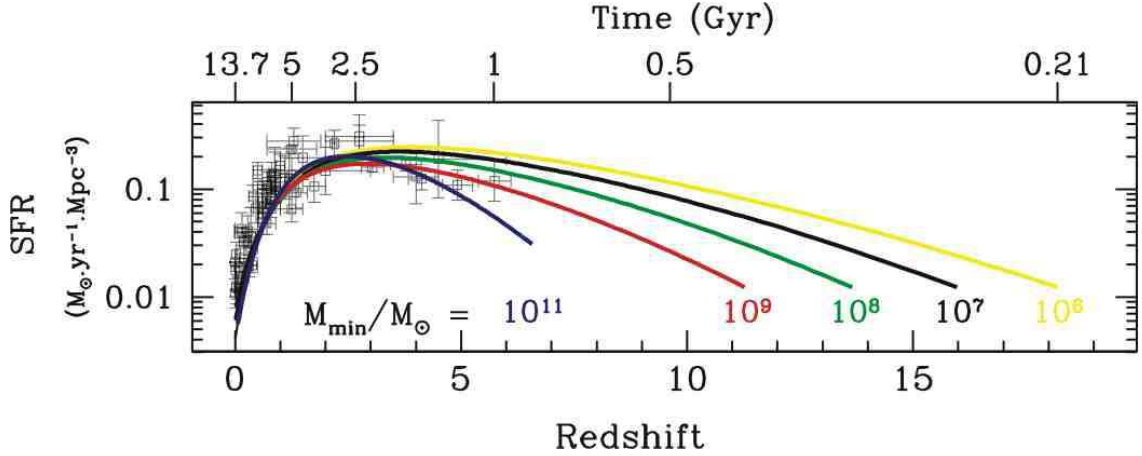,width=15.0cm}}
%\vspace*{1.4truein}		%ORIGINAL SIZE=1.6TRUEIN x 100% - 0.2TRUEIN
%\leftline{\hfill\vbox{\hrule width 5cm height0.001pt}\hfill}
\caption{The cosmic star formation rate for the normal mode of star formation (Model 0). Plotted are the results of the best fit models for each choice of $M_\mathrm{min}$.}
\label{fig:model0SFR}
\end{figure}

The results of the best fit models are plotted in Figure~\ref{fig:model0SFR} for 
each value $M_\mathrm{min}$ as indicated. Because we have fixed the initial baryon fraction in structures, each value of $M_\mathrm{min}$ corresponds to a different initial redshift. 
As one can see, each of the models gives a satisfactory fit to the global SFR, save perhaps the case
with $M_\mathrm{min} = 10^{11}$ M$_\odot$. The data shown (taken from \cite{csfr}) 
has already been corrected for extinction. 

\begin{figure}
%\vspace*{13pt}
%\leftline{\hfill\vbox{\hrule width 5cm height0.001pt}\hfill}
      \mbox{\epsfig{figure=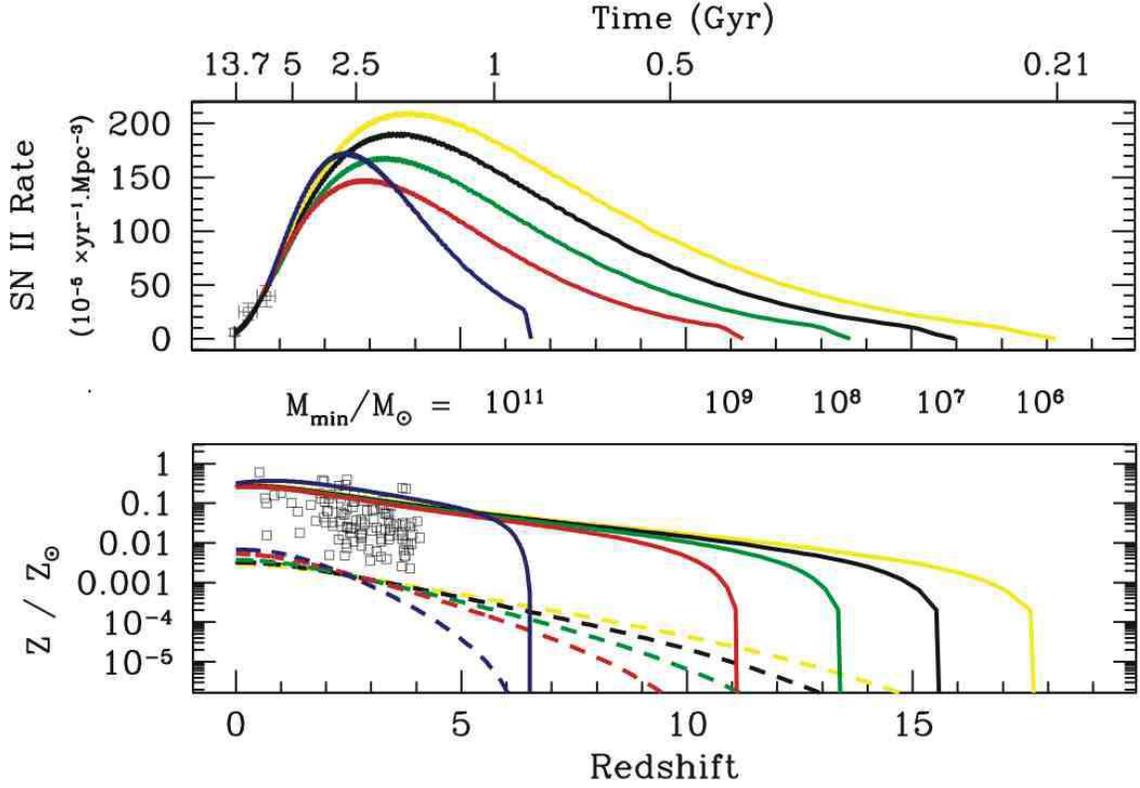,width=15.0cm}}
%\vspace*{1.4truein}		%ORIGINAL SIZE=1.6TRUEIN x 100% - 0.2TRUEIN
%\leftline{\hfill\vbox{\hrule width 5cm height0.001pt}\hfill}
\caption{The type II supernova rate and global metallicity for Model 0. Plotted are the results of the best fit models for each choice of $M_\mathrm{min}$.}
\label{fig:model0SNZ}
\end{figure}

Some of the consequences of the SFRs shown in Figure~\ref{fig:model0SFR}
are shown in Figure~\ref{fig:model0SNZ}, where the type II supernova rate and metal enrichment are shown. The observed rates of type II supernovae up to $z \sim 0.7$ is taken from \cite{dahlen}.
The predicted type II supernova rates are also an excellent fit to the 
existing data at low redshift. The predicted rates are substantially higher at higher redshift and lower
$M_\mathrm{min}$. Also shown is the evolution of the metallicity in both the ISM and IGM (dashed curves). While the ISM metallicity rises very quickly initially,  evolution is more gradual in the IGM
where it is controlled by the outflow. The metallicity of 100 Damped Lyman-$\alpha$ systems as measured by \cite{prochaska} is also plotted. Note that only the normal mode is included here,
and contributions from Population III stars will only serve to increase the metallicity in both the ISM and IGM.

\section{The massive mode}

We consider three different models, labeled Models 1, 2a, and 2b to describe the massive mode.
They are distinguished by their respective stellar mass ranges.
In Model 1,  the IMF is defined for stars with masses in the range 40 M$_{\odot} \le m \le 100$ M$_{\odot}$. 
All of these stars die in core collapse supernovae leaving a black hole remnant with the mass of the progenitor's helium core.
They all contribute to the chemical enrichment of the ISM and IGM.
This period of star formation is brief and is described by a SFR of 
the form
\beq
\psi_2=\nu_2 M_{\rm struct} e^{-Z_{{\rm IGM}}/Z_\mathrm{crit}}
\eeq 
where $M_{\rm struct}$ is the mass of the primitive star forming structure and  $Z_\mathrm{crit} = 10^{-4}$ Z$_\odot$ is the critical metallicity at which Population III star formation 
ends\cite{zcrit}.  Other parameters describing these models are given in Table 2
for the case of $M_{\rm min} = 10^7$ M$_\odot$.  

\begin{table}[ht]
\begin{center}
\begin{tabular}{|lccc|}
\multicolumn{4}{c}{\textbf{Massive mode}}  \\ %& \multicolumn{2}{|c}{\textbf{Massive mode}}\\
\hline
Model & $M_{\mathrm{min}}$ & 
$\epsilon$ & 
$\nu_{2}$  \\
& (M$_{\odot}$) 
 & 
 & 
(Gyr$^{-1}$)  \\
\hline
\hline
1 & $10^{7}$  &  $2 \times 10^{-3}$ & 60 \\
\hline
1e & $10^{7}$  &  $6 \times 10^{-5}$ & 340 \\
\hline
2a & $10^{7}$  &  $1 \times 10^{-3}$ & 9 \\
\hline
2ae & $10^{7}$  &  $8 \times 10^{-5}$ & 40 \\
\hline
2b & $10^{7}$  &  $3 \times 10^{-3}$ & 100 \\
\hline
\end{tabular}
\end{center}
\caption{Parameter values for the massive starburst Models 1, 2a and 2b. Column 1 indicates the 
model number and column 2 the input value of the minimum mass for star forming structures.   In columns 3 and 4, we show the adopted outflow efficiency and the coefficient of the massive mode SFR.}
\label{tab:model1}
\end{table}

When a massive mode is added to the normal mode described by Model 0,
the outflow efficiency must be adjusted so as to avoid the overproduction of 
metals in the IGM.  However, there is a degeneracy in the massive mode parameters
$\epsilon$ and $\nu_2$.  In the model labeled 1, the massive mode
contributes roughly 50\% of the IGM metallicity at a redshift $z = 2.5$.
By increasing $\nu_2$ and decreasing $\epsilon$, this contribution can
be increased to 90\%, at the same time increasing the ionization capacity of the model.
This case is labeled 1e.

\begin{figure}[ht]
%\vspace*{13pt}
%\leftline{\hfill\vbox{\hrule width 5cm height0.001pt}\hfill}
      \mbox{\epsfig{figure=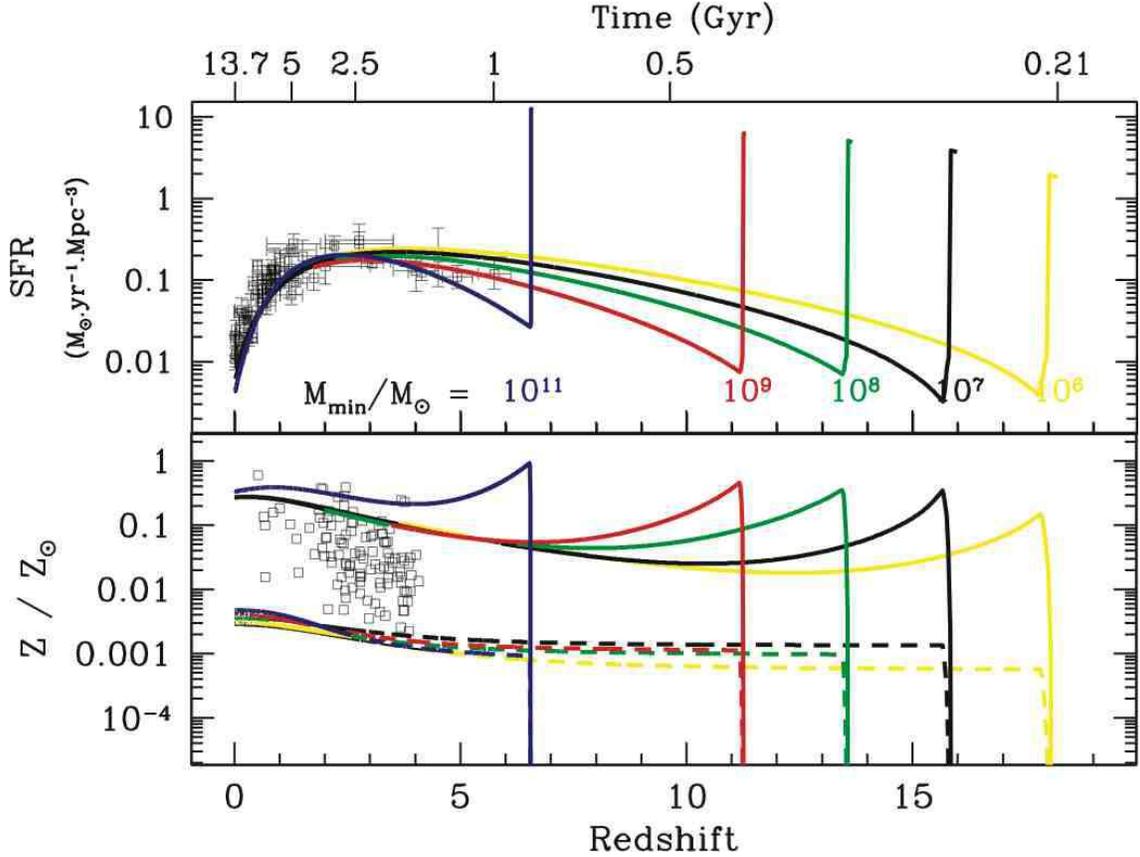,width=15.0cm}}
%\vspace*{1.4truein}		%ORIGINAL SIZE=1.6TRUEIN x 100% - 0.2TRUEIN
%\leftline{\hfill\vbox{\hrule width 5cm height0.001pt}\hfill}
\caption{The cosmic star formation rate and global metallicity for the normal mode with an added burst of massive stars (Model 1).}
\label{fig:model1SFRZ}
\end{figure}

As seen in Figure~\ref{fig:model1SFRZ}, star formation in Model 1
begins at a very high rate and falls precipitously as metals are injected into the ISM. 
Below we will also consider a variant of Model 1 in which the massive mode begins much earlier
but its duration is extended in redshift.
As in the case of Model 0, the onset of star formation is determined by $M_\mathrm{min}$ and the initial value for the baryon fraction in structures (fixed to be 1\%). 
As one can see in the lower panel, the metallicity in the ISM reaches 
values far in excess of $Z_\mathrm{crit}$ due to the finite lifetime of the massive
stars relative to the speed at which the metallicity is attained. Notice also that
once the metallicity from Pop III stars is produced, the ISM metallicity later decreases as a result of
the accretion of metal-free gas as the structures grow.

\begin{figure}[ht!]
%\vspace*{13pt}
%\leftline{\hfill\vbox{\hrule width 5cm height0.001pt}\hfill}
      \mbox{\epsfig{figure=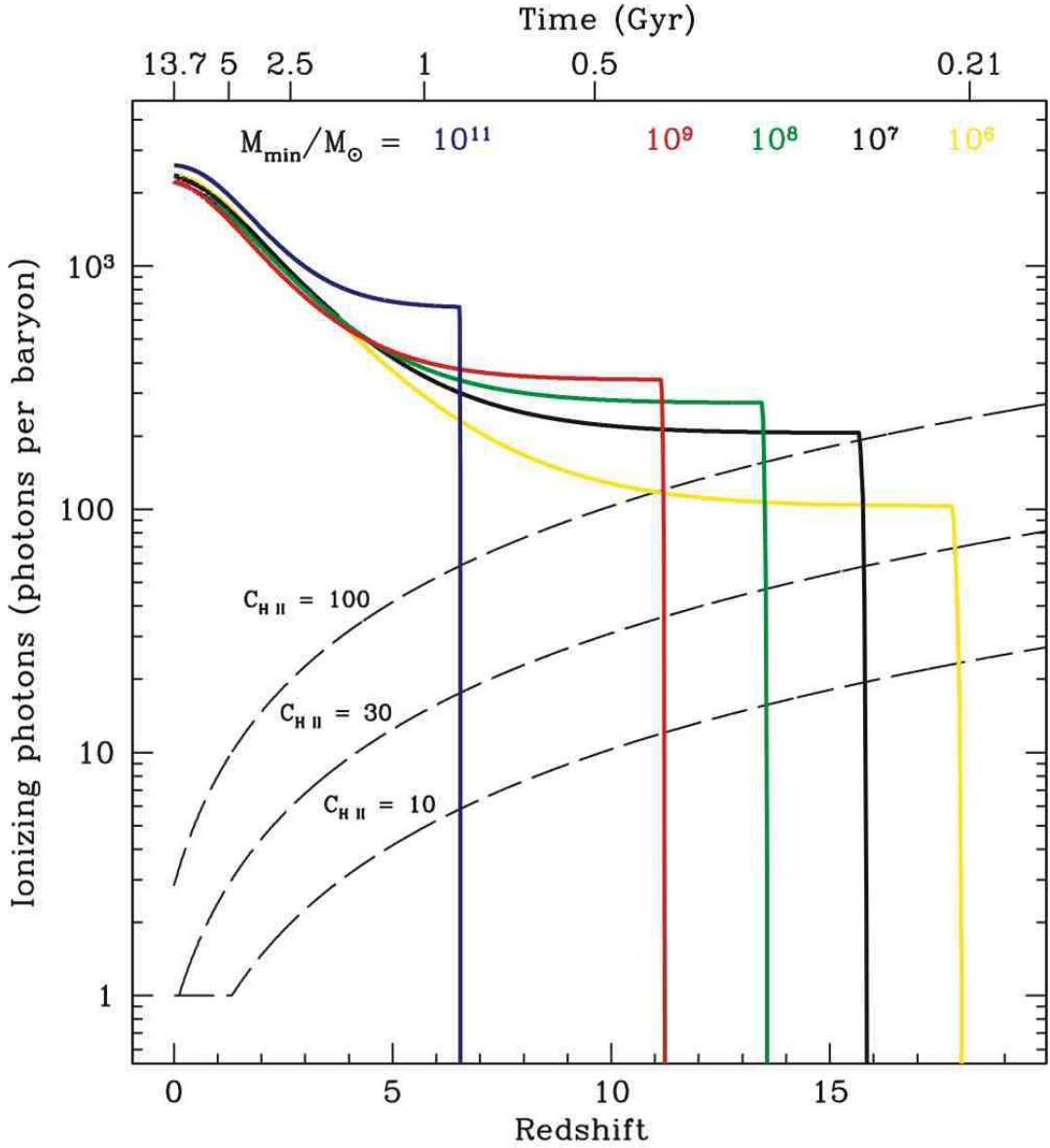,width=15.0cm}}
%\vspace*{1.4truein}		%ORIGINAL SIZE=1.6TRUEIN x 100% - 0.2TRUEIN
%\leftline{\hfill\vbox{\hrule width 5cm height0.001pt}\hfill}
\caption{The cumulative number of ionizing photons emitted by stars is plotted as a function of redshift for all models of Figure~\protect{\ref{fig:model1SFRZ}}. The minimum number of photons per baryon necessary to fully reionize the IGM is plotted as a function of redshift by thin dashed lines for three different values of the clumpiness factor $C_\mathrm{H\ II}=10$, 30 and 100.}
\label{fig:model1Reionization}
\end{figure}

Among the chief motivating factors in developing a model of cosmic chemical evolution is
the early reionization of the Universe.  Several studies suggest
 that an early burst of star formation,
as in Model 1, is sufficient\cite{venk,daigne1,daigne2}.
In Figure~\ref{fig:model1Reionization}, we show  the number of ionizing photons per baryon produced for each of our choices of $M_\mathrm{min}$. 
The procedure for  computing this stellar ionizing flux is explained in \cite{daigne1}. It is important to remember that only a fraction $f_\mathrm{esc}$ of these UV photons will escape the structures and therefore be available to ionize the IGM. The effective value of $f_\mathrm{esc}$ is poorly known but could vary from about 1 to 30\%.
The minimum number of photons required for complete reionization is also plotted for three possible clumpiness factors. The ionizing potential  clearly increases with $M_\mathrm{min}$ and decreasing redshift. The ratio of this minimum number (dashed line) to that for the stellar ionizing photons (solid line) gives the minimum fraction $f_\mathrm{esc}$ necessary to fully reionize the IGM.

\begin{figure}[ht!]
%\vspace*{13pt}
%\leftline{\hfill\vbox{\hrule width 5cm height0.001pt}\hfill}
      \mbox{\epsfig{figure=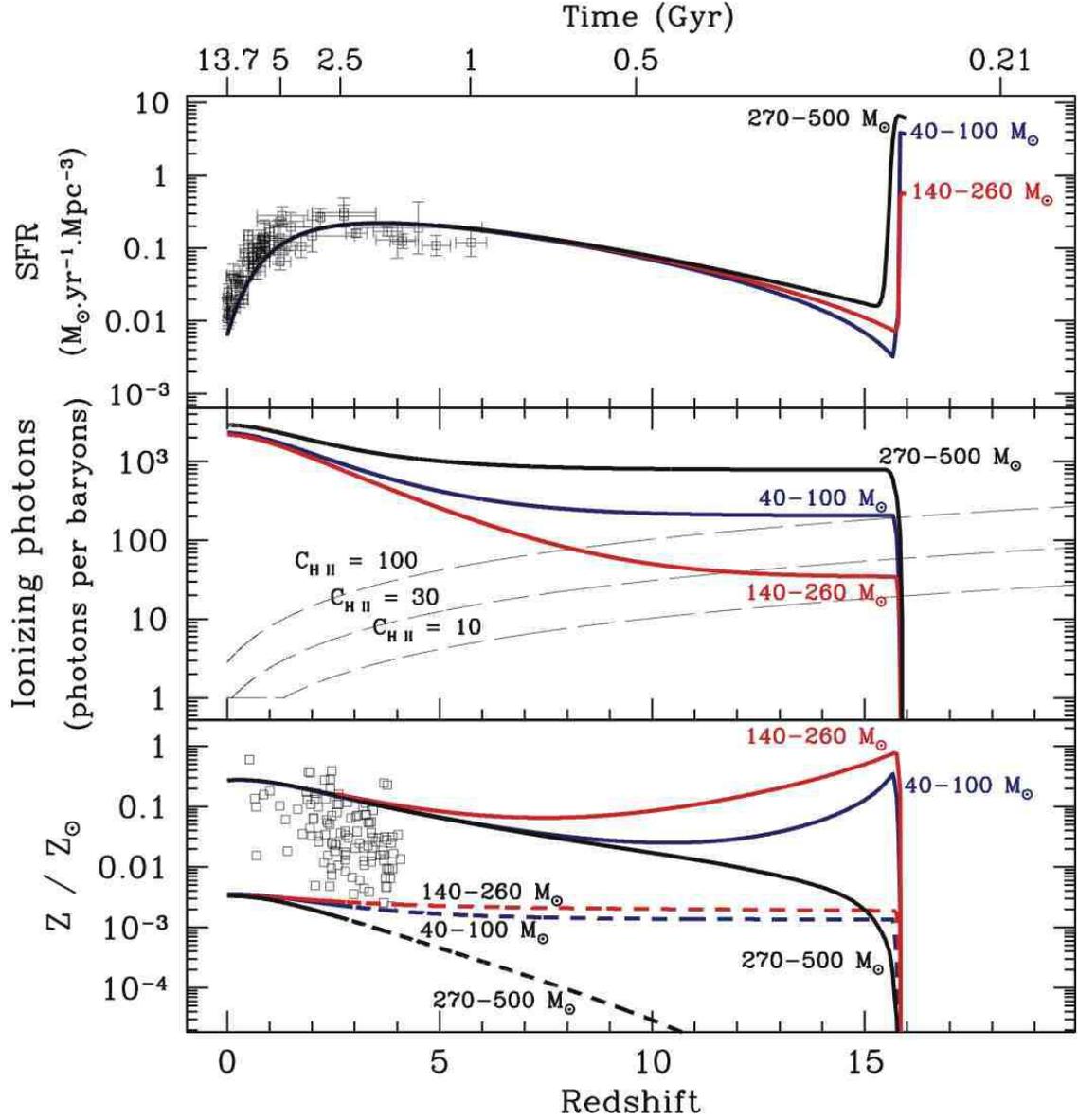,width=15.0cm}}
%\vspace*{1.4truein}		%ORIGINAL SIZE=1.6TRUEIN x 100% - 0.2TRUEIN
%\leftline{\hfill\vbox{\hrule width 5cm height0.001pt}\hfill}
\caption{The star formation rate, reionization potential, and global metallicity for Models 1, 2a, and 2b with $M_\mathrm{min}=10^{7}\ \mathrm{M_{\odot}}$. }
\label{fig:model2SFRZ}
\end{figure}

Model 2a is described by
very massive stars which become pair instability supernovae.  The IMF is defined for 140 M$_{\odot} \le m \le 260$ M$_{\odot}$ and the SFR for this model is the same as that for Model 1, but must be reduced by a factor of 8 due to constraints on metal abundances in the ISM. 
The most massive stars are considered in Model 2b and fall in the range 270 M$_{\odot} \le m \le 500$ M$_{\odot}$, with the SFR as in Model 1. These stars entirely collapse into black holes and do not contribute 
to the chemical enrichment of either the ISM or IGM.  A comparison of the star formation rates,
ionization potentials, and global metallicity in models 1, 2a, and 2b is shown in 
Figure~\ref{fig:model2SFRZ}.

Rather than the rapid burst models for Population III stars discussed above,
it is useful to consider the consequences of an extended period of 
Population III star formation as described in \cite{daigne2}. 
The non-burst models begin at high redshift, $z \approx 30$, independent of the
the minimum halo mass. As such, they do not depend on any choice of the initial
baryon fraction.  In this case, we adopt a SFR given by
\begin{equation}
\psi_2 = \nu_{2} M_{\rm ISM} e^{-Z_{{\rm IGM}}/Z_\mathrm{crit}}
\end{equation}
Notice the replacement of the structure mass with the
mass of gas in the structure ISM.

Given the new SFR, it is necessary to refit the observed data and once again maximize
$\nu_2$ for ionization without  the overproduction of metals.  New parameter values of 
outflow and the SFR are given in Table \ref{tab:model1n} for non-burst Models 1, 2a, and 2b,
now labeled 1n, 2an, and 2bn respectively.

\begin{table}[ht]
\begin{center}
\begin{tabular}{|lccc|}
\multicolumn{4}{c}{\textbf{Massive mode}}  \\ %& \multicolumn{2}{|c}{\textbf{Massive mode}}\\
\hline
Model & $M_{\mathrm{min}}$ & 
$\epsilon$ & 
$\nu_{2}$  \\
& (M$_{\odot}$) 
 & 
 & 
(Gyr$^{-1}$)  \\
\hline
\hline
1n & $10^{7}$  &  $6 \times 10^{-4}$ & 80 \\
\hline
2an & $10^{7}$  &  $1 \times 10^{-4}$ & 40 \\
\hline
2bn & $10^{7}$  &  $5 \times 10^{-3}$ & 10 \\
\hline
\end{tabular}
\end{center}
\caption{As in Table \protect\ref{tab:model1} parameter values for  
nonburst Models 1n, 2an and 2bn. }
\label{tab:model1n}
\end{table}

\begin{figure}
\begin{center}
\resizebox{\textwidth}{!}{\includegraphics{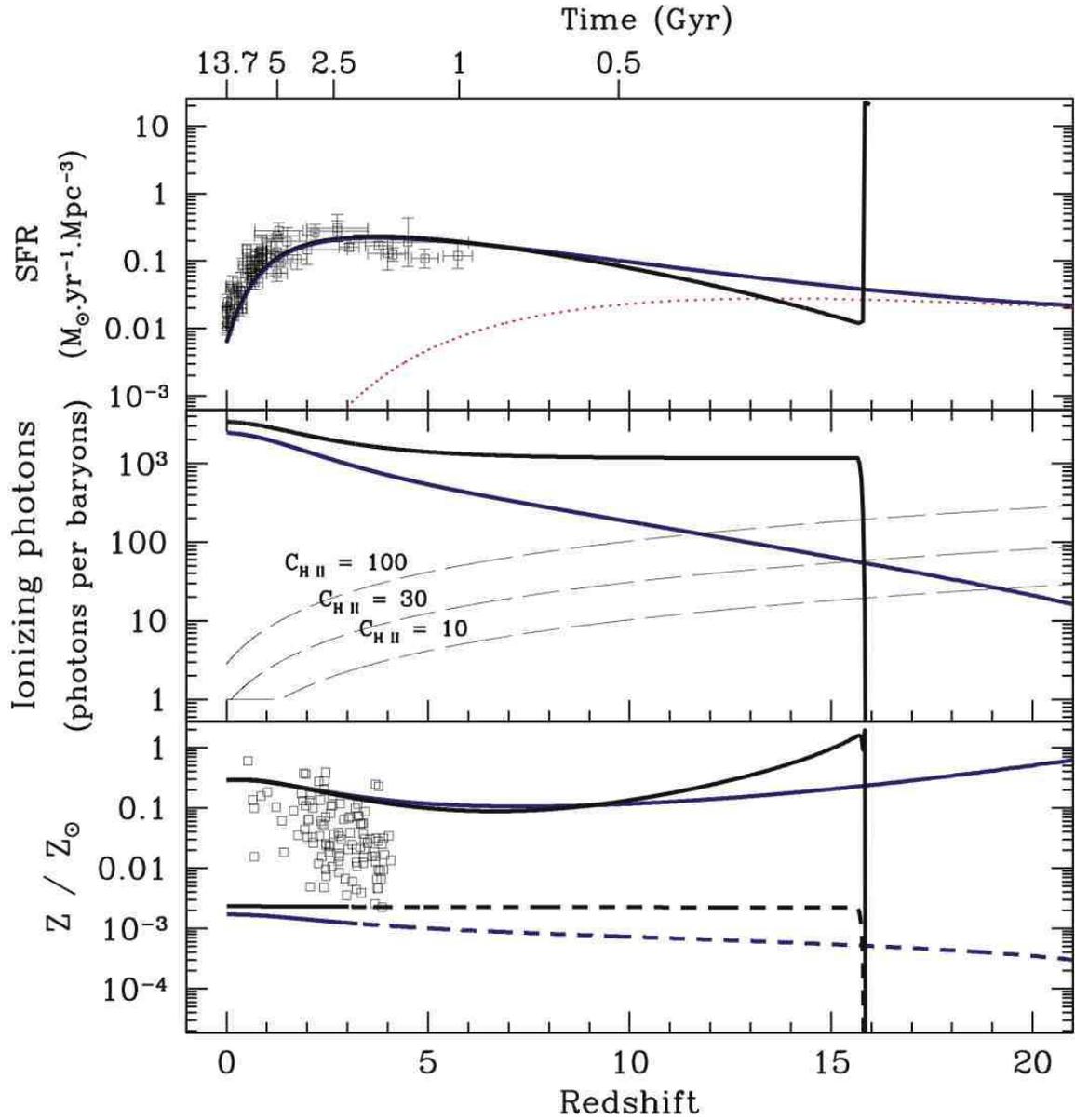}}
\end{center}
\caption{The star formation rate, ionizing photons per baryon and global metallicity for Model 1n and a rapid burst Model 1e with $M_\mathrm{min}=10^{7}\ \mathrm{M_{\odot}}$.
The dotted line in the upper panel shows the SFR of the massive mode of Model 1n.}
\label{fig:extremModel1SFRFracBZ}
\end{figure}

In Figure~\ref{fig:extremModel1SFRFracBZ}, 
we show the SFR, ionization potential, and metallicity for such a non-burst model
compared with the analogous result for Model 1e, restricting attention to  
$M_\mathrm{min} = 10^7$ M$_\odot$.  In this case, massive star formation begins at $z = 30$.
The dotted curve in the top panel of Figure~\ref{fig:extremModel1SFRFracBZ} is the 
massive component of Model 1n.  While the massive mode SFR in Model 1n is significantly lower than
the peak burst of 1e, its duration is far longer and continues down to redshifts $ < 3$,
although the total star formation rate becomes dominated by the normal mode
when $z \la 10$. 
The clear difference between Model 1n and the rapid burst Model 1e  is the production of
ionizing photons at high redshift as shown in the middle panel. 
The burst is capable of ionizing the IGM at $z = 16$ for all three 
choices of $C_\mathrm{H\ II}$.
  In contrast, the metallicity shown in the lower panel is relatively unaffected by
  the rapid burst.

Finally, given an IMF and SFR, it is straightforward to compute the
rate of core collapse supernovae,
\begin{equation}
{\mathrm SNR} =\int_{max(8\mathrm{M_{\odot}}, m_\mathrm{min}(t))}^{m_\mathrm{sup}} 
dm\ \phi(m)\psi\left(t-\tau(m)\right) \ ,
\end{equation}
where $m_\mathrm{min}(t)$ is the minimum mass with lifetime, $\tau$, less than $t$.

\section{Neutrino Production}

In all of the models we will consider, star formation begins at high redshift, dominated 
initially by massive stars which may explode as core collapse or pair-instability supernovae and
provide for the reionization of the IGM. Each explosion, regardless of type, releases most
of the star's gravitational energy in the form of neutrinos with a specified energy spectrum and flux.
Given a chemical evolutionary model, the integrated contribution of SN to the neutrino background can be computed.

The expected differential flux of neutrinos at Earth with energy $E$ can be expressed as 
%\beq
 %\frac{dF_\alpha}{dE} =  \int_{M_{min}}^{M_{max}} \int_{z_f}^{z_i} N_{\nu_\alpha} \, \phi(m) \, \psi \, (1+z) \left| \frac{dt}{dz} \right| \frac{dP_\alpha}{dE} \,\, dm \, dz 
%\label{diffnu}
%\eeq
\beq
\frac{dF_\alpha}{dE} =
\int_{0}^{z_i} dz\, (1+z) \left| \frac{dt}{dz} \right|
\int_{M_{min}}^{M_{max}} dm\, \phi(m) \, \psi\left(t-\tau(m)\right) \,
N_{\nu_\alpha}(m)\, \frac{dP_\alpha}{dE'}
\label{diffnu}
\eeq
where $N_{\nu_\alpha}(m)$ is the total number of neutrinos of a given species, $\alpha$, emitted in the core collapse of a star of mass $m$,  and $\frac{dP_\alpha}{dE'}$ represents the neutrino energy spectra at emission with energy $E' = E(1+z)$. The integration limits $M_{min}$ and $M_{max}$ are the minimum and maximum masses in each model for which supernovae occur, and $z_i$ is the initial redshift at which star formation begins. 

When a star undergoes core collapse, the mass of the remnant is determined by the mass of the progenitor.  We assume that all stars of mass $ m \ga 8$ M$_{\odot} $ will die as supernovae. For stars of mass 8 M$_{\odot} < m < 30$ M$_{\odot} $, the remnant after core collapse will be a neutron star of $ m \approx 1.5$~M$_{\odot} $. More massive stars fall into two categories; black holes and pair instability supernovae. Pair instability supernovae are thought to occur for stars with 140~M$_{\odot} \la m \la 260$ M$_{\odot} $, in which case the explosion leaves no remnant. All other stars collapse to form black holes. Stars with 30 M$_{\odot}<m<100$ M$_{\odot} $ become black holes with mass approximately that of the star's helium core before collapse\cite{stardeath}. We take the mass of the Helium core to be 
\beq
M_{He}=\frac{13}{24} \cdot (m-20\,M_{\odot}) 
\eeq
for a star with main sequence mass $m$~\cite{ww2002}. We assume that stars with $ m>260$ M$_{\odot} $ collapse entirely to black holes.

The energy emitted in each core collapse, $E_{cc}$ corresponds to the change in gravitational energy, 99\% of which is emitted as neutrinos\cite{by}. In the cases where collapse results in a neutron star, $E_{cc}=5 \times 10^{53}\,$ergs. For stars that collapse to black holes, $E_{cc}$ is proportional to the mass of the black hole. For masses less than 100 M$_\odot$, we take $E_{cc} =  0.3M_{He}$ . Pair instability supernovae experience a much more powerful explosion than core collapse supernovae, however few neutrinos are emitted and with very low energies such that they would not be observed\cite{wwm1986}.
In this case, we assume the energy emitted in neutrinos is the same as that for ordinary core collapse supernovae yielding a neutron star, but the average neutrino energy is $\langle E_{\bar{\nu}_e} \rangle = 1.2 \, \textrm{MeV} $~\cite{wwm1986}.
For the most massive stars considered, $E_{cc} = 0.3 m$. Although several studies find a distinct hierarchy in the partitioning of neutrino luminosity among the species during the different luminosity phases of core collapse, equipartition of the total energy emitted by the star is generally accepted\cite{tot,as,sksw}. For a comparison of luminosity hierarchies found in recent simulations, see Keil et al.\cite{krj}. 

We assume that each electron neutrino carries an average energy $\langle E_{\nu_e}\rangle = 13.3 \,\textrm{MeV}$, which is consistent with simulations. The charged current reactions that prevent neutrinos from emerging from the star are $\nu
_{\textrm{e}} \textrm{n} \to \textrm{p} \textrm{e}^-$ and $\bar{\nu}_\textrm{e}\textrm{p} \to \textrm{n} \textrm{e}^+$. The different trapping reactions result in different neutrinosphere radii, and therefore different average energies for $\nu_e$ and $\bar{\nu}_e$. We assume $\langle E_{\bar{\nu}_e}\rangle = 15.3 \,\textrm{MeV}$. The other species, denoted $\nu_x$, undergo only neutral current interactions.  The mechanism that governs their average temperature at emission is more complicated, but the generally accepted hierarchy is $\langle E_{\nu_e}\rangle < \langle E_{\bar{\nu}_e}\rangle < \langle E_{\nu_x}\rangle$. 
We have taken $\langle E_{\nu_x} \rangle = 20$ MeV.
The total number of $\nu_{\alpha}$ emitted by a star during core collapse is given by 
\beq
N_{\nu_\alpha}=\frac{E_{cc}}{\langle E_{\nu_\alpha}\rangle}.
\eeq

The neutrino spectra at emission can be described by a normalized Fermi-Dirac distribution, 
%\beq
 %\frac{dP_{\alpha}}{dE}=\frac{2}{3 \zeta_3 T_{\alpha}^3} \frac{E^2(1+z)^2}{e^{E(1+z)/T_{\alpha}}+1} 
 %\eeq 
 \beq
  \frac{dP_{\alpha}}{dE'}=\frac{2}{3 \zeta_3 T_{\alpha}^3} \frac{{E'}^2}{e^{E'/T_{\alpha}}+1}
\eeq
 where $T_{\alpha}= {180 \zeta_3 \langle E_{\nu_\alpha}\rangle}/{7 \pi^4}$ is the effective neutrino temperature taken to be independent of the mass of the star.
%  and $E$ is the redshifted neutrino energy at detection. 
 We assume a flat $\Lambda$CDM cosmology with  
 \beq
 \left| \frac{dt}{dz} \right| = \frac{9.78\,h^{-1}\,\textrm{Gyr}}{(1+z)\sqrt{\Omega_{\Lambda}+\Omega_m(1+z)^3}}
 \eeq
 where $\Omega_{\Lambda}=0.73$, $\Omega_m=0.27$, and $h=0.71$~\cite{wmap}.

\section{Resulting Neutrino Fluxes}

\begin{figure}[ht]
\centering
\includegraphics[width=0.58\textwidth]{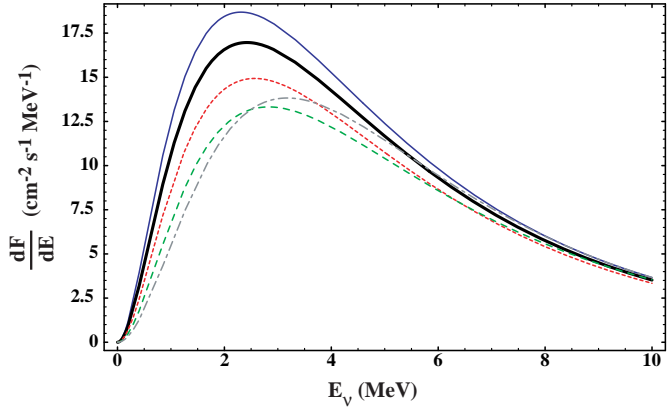}
\caption{Fluxes from Model 0 for five choices of $M_\mathrm{min} = 10^6$ (solid), $10^7$ (thick), $10^8$ (dotted), $10^9$ (dashed) and 
$10^{11}$ (dot-dashed) M$_\odot$.}
\label{model0s2}
\end{figure}

Having described the chemical evolution models and the calculation of the
neutrino spectrum, we now show the results for each of the models considered\cite{ourpaper}.
Model 0 fluxes are plotted in Figure~\ref{model0s2}. Results are shown for several choices of minimum halo masses, $M_\mathrm{min}$. As we will see below these fluxes are large enough to be probed by
current detectors. As $M_\mathrm{min}$ is increased, star formation occurs at later redshift and as a 
result, the peak of the neutrino flux is shifted slightly to higher energy.

\begin{figure}[ht]
\centering
\includegraphics[width=0.58\textwidth]{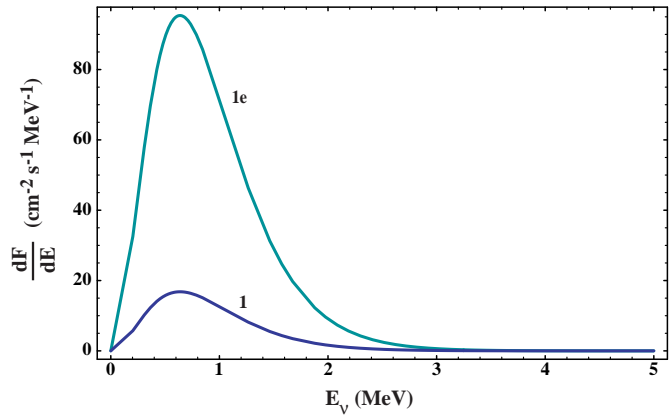}
\caption{Fluxes from the massive modes of Model 1 (black) and 1e (grey).}
\label{model1s2}
\end{figure}

The massive modes of Model 1 and 1e fluxes are plotted in Figure~\ref{model1s2} for the specific choice of $M_\mathrm{min} = 10^7$ M$_\odot$ which is the preferred case in \cite{daigne2}. As seen in Figure~\ref{fig:model2SFRZ}, massive stars associated with Population III turn on
at a redshift of approximately 16, but the duration of the burst is relatively brief.
As a result, the peak of the flux distribution is at relatively low energy. More importantly,
because of the brevity of the burst, the {\em entire} neutrino spectrum is redshifted down,
in contrast to the Model 0 spectrum which extends to higher energy due to stars produced at lower
redshifts.
As expected, the more extreme model, 1e, has a peak flux which is about 5 times that found for Model 1.
This is directly related to the increased SFR in Model 1e as characterized by the increase in $\nu_2$.

\begin{figure}[ht]
\centering
\includegraphics[width=0.58\textwidth]{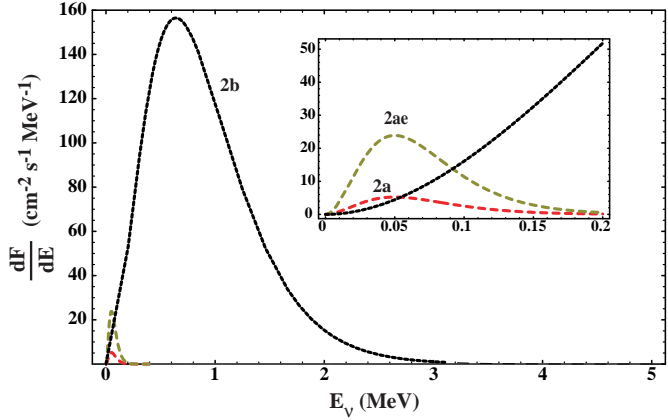}
\caption{Fluxes from the massive modes of Model 2a and 2ae (dashed), and 2b (dotted).
\label{model2abs2}}
\end{figure}

\begin{figure}[ht]
\centering
\includegraphics[width=0.58\textwidth]{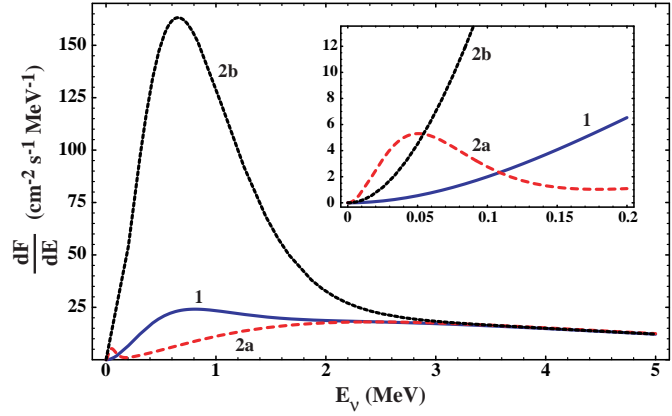}
\caption{Total fluxes for Model 1 (solid), 2a (dashed), and 2b (dotted).
\label{totalfluxs2}}
\end{figure}

Similarly, we show in Figure~\ref{model2abs2} the resulting flux from very massive Population III stars
corresponding to Models 2a, 2ae and 2b\footnote{Since stars associated with Model 2b do
not contribute to element enrichment, there is no Model 2be.}.  
As before, the fluxes from Models 2a are relatively small
and peak at very low energy as seen in the insert to the figure.
In Figure~\ref{totalfluxs2}, we show the total fluxes in Models 1, 2a, and 2b 
with $M_\mathrm{min} = 10^7$~M$_\odot$.  
As one expects, the low energy spectrum is 
dominated by neutrinos produced in the massive mode, whereas the spectrum at
higher energies ($E_\nu \ga 3$ MeV), is indistinguishable between the models and dominated
by the normal mode.

\begin{figure}[ht]
\centering
\includegraphics[width=0.58\textwidth]{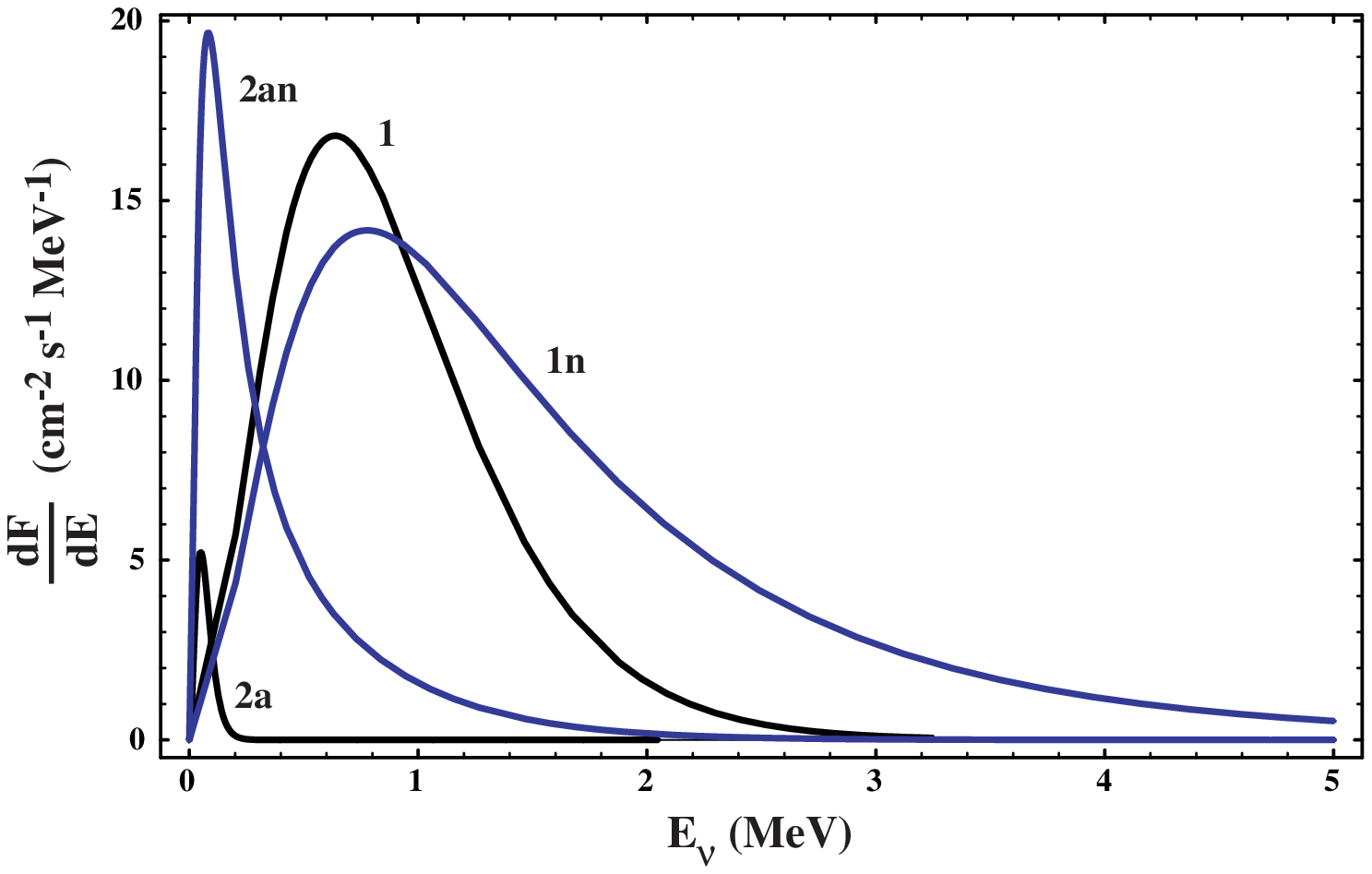}
\caption{Fluxes from the massive modes of Models 1n  and 2an (grey) and 1 and 2a (black).}
\label{model12n}
\end{figure}

\begin{figure}[ht]
\centering
\includegraphics[width=0.58\textwidth]{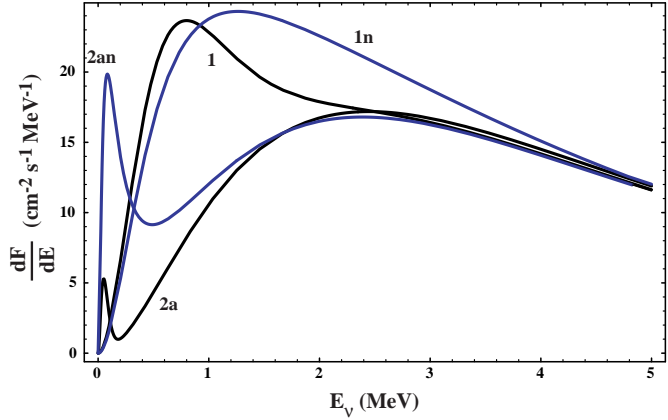}
\caption{Total fluxes from Models 1n and 2an (grey) and 1 and 2a (black).}
\label{model12nt}
\end{figure}

\begin{figure}[ht]
\centering
\includegraphics[width=0.58\textwidth]{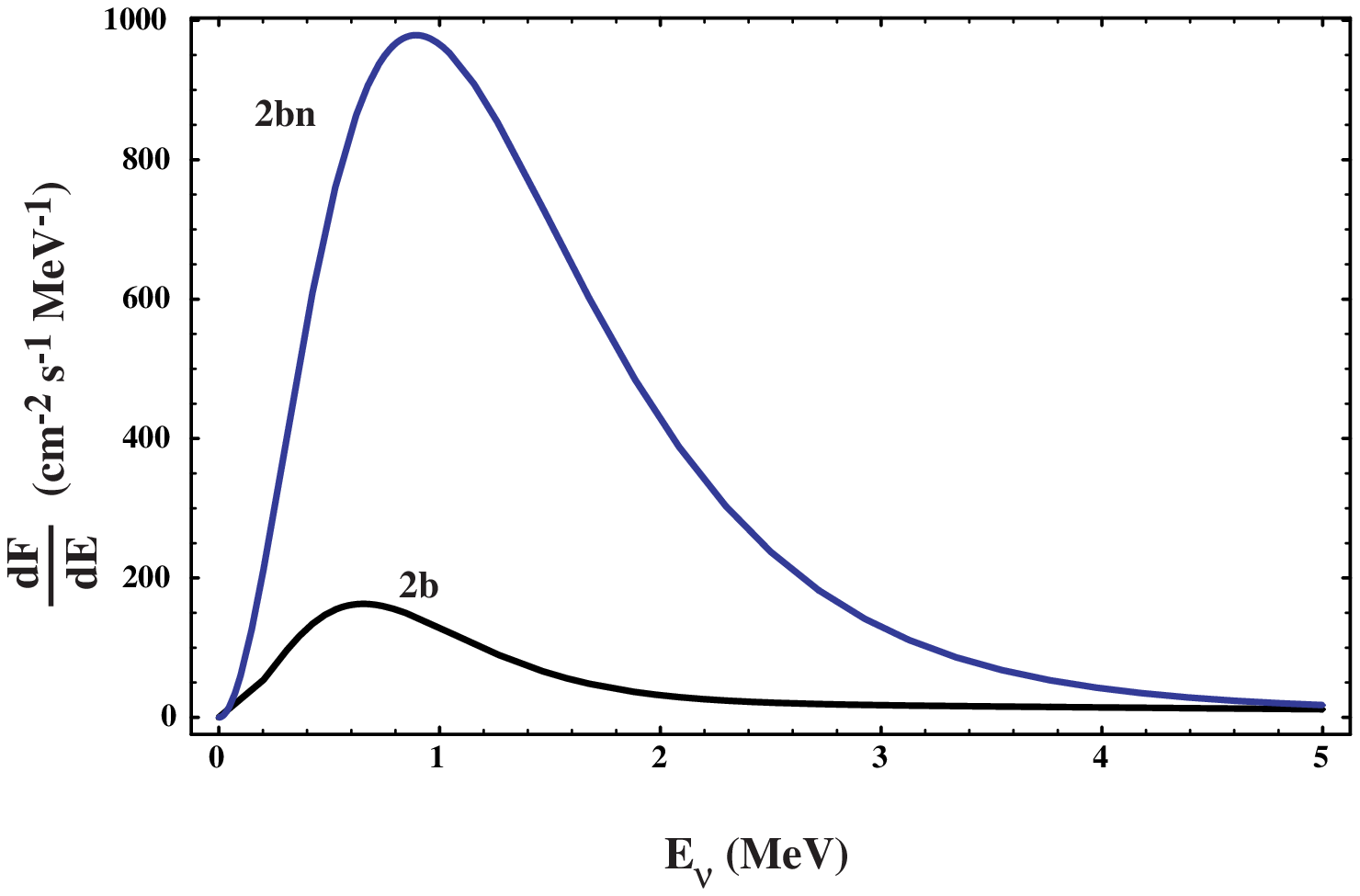}
\caption{Total fluxes from Models 2bn (grey) and 2b (black).}
\label{model2bn}
\end{figure}

Finally, we show the results for the non-burst Models 1n, 2an, and 2bn as compared with the
burst models described above.  In Figure~\ref{model12n}, we show 
the fluxes from the massive modes of models 1n and 2an compared with models 1 and 2a.
As can be expected, the fluxes extend to higher frequencies due to the longer duration of the
stellar population. The total fluxes including the normal mode are shown in Figure~\ref{model12nt}.
Similarly, in Figure~\ref{model2bn} we compare the fluxes of Models 2b and 2bn.
Because the fluxes are so large, they dominate the normal mode flux (over the energy range shown)
and we show only the total flux for this case.

\section{Detectability}

The solar neutrino flux at Earth is larger than the expected flux from SRN's by several orders of magnitude for $E_{\nu} \la 19\,$MeV~\cite{as}. However, neutrinos, rather than antineutrinos, are produced in the thermonuclear reactions in the sun and have a smaller cross section for detection by about two orders of magnitude. This, with the directional information from recoil electrons in the detector, allows this background to be excluded at SuperK, KamLAND, and SNO. But decays of spalled nuclei from cosmic ray muons constitute an unavoidable background in this range.
The current upper limit for the flux of SRN's at SuperK is $1.2\,$cm$^{-1}$s$^{-1}$ for $E_{\nu}>19.3\,$MeV~\cite{SKlimit}.
In addition, it has been pointed out recently \cite{sno}  that if the background analysis from SK is coupled with the sensitivity to electron neutrinos at SNO it will be possible to reduce the upper limit on the flux of electron neutrinos. SNO should be sensitive to a flux of $6\,$cm$^{-2}$s$^{-1}$ in the range $22.5\,$MeV$<E_{\nu_e}<32.5\,$MeV. 

\begin{figure}[ht]
\centering
\includegraphics[width=0.58\textwidth]{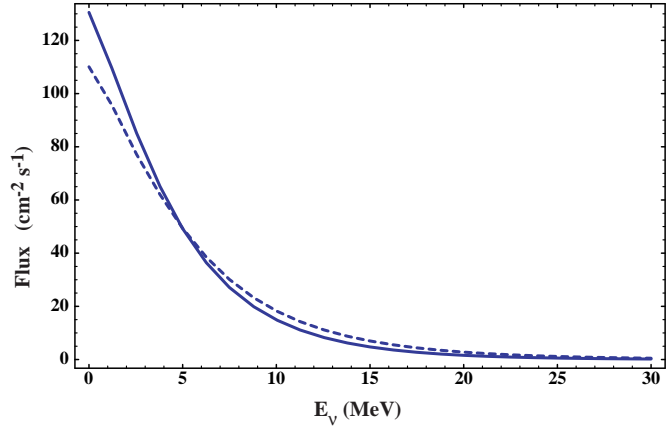}
\caption{Detectable fluxes from Model 1 with (dashed) and without (solid) oscillations as a function of neutrino energy threshold.\label{threshold}}
\end{figure}

In Figure~\ref{threshold}, we show the observable flux 
\beq
F(E_\mathrm{thresh}) = \int^\infty_{E_\mathrm{thresh}} {dF \over dE} dE
\eeq
as a function of detector threshold energy. 
While the fluxes are quite appreciable at low threshold energies, they in fact remain
relatively high at larger energies due to the large SFR associated with Model 0.
Indeed, in all of our models, our predicted flux above 19.3 MeV already exceeds
the current bound of  $1.2$ cm$^{-1}$s$^{-1}$ from SuperK~\cite{SKlimit}. The detailed flux
predictions are given in 
Table~\ref{numberflux}, where we show the detectable flux for the viable energy windows at SK and SNO for both the burst and non-burst models. Although SRN's will likely not be seen at SNO given these flux levels, in many of our models the SK bound is saturated by the expected flux, indicating that SRN's may be observed in the near future.
Models labeled osc, include the effects of neutrino oscillations.  

\begin{table}[ht]
\begin{center}
\begin{tabular}{|l|c|c|}
\hline
Model & SK Flux & SNO Flux \\
\hline
\hline
0 & 1.8 & 0.47 \\
0n & 1.8 & 0.46 \\
1 & 1.8 & 0.47 \\
1osc & 3.2 & 1.4 \\
1e & 1.9 & 0.49 \\
1n & 1.8 & 0.47 \\
2a & 1.9 & 0.48 \\
2ae & 1.9 & 0.49 \\
2an & 1.8 & 0.47 \\
2b & 1.8 & 0.47 \\
2bosc & 3.2 & 1.4 \\
2bn & 1.7 & 0.44 \\
\hline
\end{tabular} 
\caption{Predicted fluxes in cm$^{-2}$s$^{-1}$ in the models considered here. Results are given for electron antineutrinos with energies $E_{\bar{\nu}_e}>19.3$MeV for SK and for electron neutrinos with $22.5$MeV$<E_{\nu_e}<32.5$MeV for SNO. \label{numberflux}}
\end{center}
\end{table}

\begin{figure}[ht]
\centering
\includegraphics[width=0.58\textwidth]{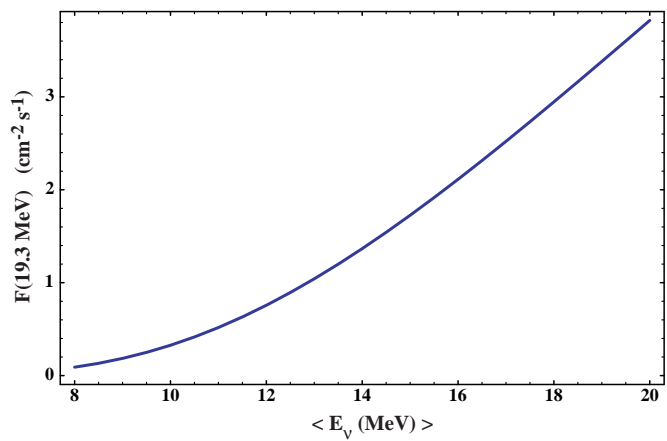}
\caption{The integrated flux above 19.3 MeV in Model 1 as a function of the
average neutrino energy.
\label{aveE}}
\end{figure}

Despite the large fluxes displayed in Table \ref{numberflux} relative to the SK limit\cite{SKlimit}, 
one can not conclude that the models considered have already been excluded by experiment.
There are of course many uncertainties built into our chemical evolution models as well as uncertainties
in the adopted neutrino physics.  These fluxes are very sensitive to our assumed average neutrino energy.  Recall our adopted value for $E_{\bar \nu_e}$ is 15.3~MeV.  In Fig. \ref{aveE}, we show the sensitivity of the flux above 19.3 MeV to the average neutrino energy.
In order to satisfy the SK limit of 1.2 cm$^{-2}$s$^{-1}$, we would have to lower
$\langle E_{\bar \nu_e} \rangle$ to 13.3 MeV.  This is fully consistent with the range of neutrino
energies in supernova models \cite{krj}.

\section{Summary}

We have considered several scenarios for star formation which reproduce the observed chemical abundances and SFR for $z \leq 6$ and reionize the universe at high redshift. Each model of star formation here consists of a normal mode coupled to a  Population III mode of massive star formation at high redshift. We considered both a burst and non-burst model for Population III star formation.

In the burst model, 
because the massive mode of star formation is so brief and takes place at high redshift, the corresponding electron anti-neutrino fluxes peak at $E_{\nu} \la 1\,$MeV. 
Even in the non-burst model, fluxes are become small at energies above $\sim 4$ MeV.
Thus despite the large fluxes produced by the massive mode, these low energy neutrinos
will be difficult to detect. In contrast, 
the normal mode of star formation, which dominates the flux at observable energies, is peaked at a somewhat higher energy and has a broad spectrum due to the production of stars at lower redshift. 
Our calculated fluxes of SRN's from core collapse, however, saturate the SK bound of $1.2\,$cm$^{-2}$s\,$^{-1}$ for $E_{\nu}>19.3\,$MeV in all  models. Although there are uncertainties in the neutrino physics, such as the average energies at emission, the prospects for observation in the near future are good.

\section{Acknowledgements}
We thank Fr\'ed\'eric Daigne and Elisabeth Vangioni for their work on the star formation models and for useful discussions.
This work was supported in part by DOE grant
DE--FG02--94ER--40823.

\end{document}